\input{aipcheck}

\documentclass[
    ,final            % use final for the camera ready runs
  ]
  {aipproc}

\layoutstyle{8x11double}

\usepackage{amsmath}
\newcommand{\bs}[1]{\mathbf{{#1}}}

\begin{document}

\title{New Developments in MadGraph/MadEvent}


\author{J.\ Alwall}{
  address={SLAC, Stanford University, Menlo Park, CA 94025,
  E-mail: alwall@slac.stanford.edu}
}
\author{P.\ Artoisenet}{
  address={Université Catholique de Louvain, Chemin du Cyclotron
  2, B-1348 Louvain-la-Neuve, Belgium}
}
\author{S.\ {de Visscher}}{
  address={Université Catholique de Louvain, Chemin du Cyclotron
  2, B-1348 Louvain-la-Neuve, Belgium}
}
\author{C.\ Duhr}{
  address={Université Catholique de Louvain, Chemin du Cyclotron
  2, B-1348 Louvain-la-Neuve, Belgium}
}
\author{R.\ Frederix}{
  address={Université Catholique de Louvain, Chemin du Cyclotron
  2, B-1348 Louvain-la-Neuve, Belgium}
}
\author{M.\ Herquet}{
  address={Université Catholique de Louvain, Chemin du Cyclotron
  2, B-1348 Louvain-la-Neuve, Belgium}
}
\author{O.\ Mattelaer}{
  address={Université Catholique de Louvain, Chemin du Cyclotron
  2, B-1348 Louvain-la-Neuve, Belgium}
}

\begin{abstract}
We here present some recent developments of MadGraph/MadEvent since
the latest published version, 4.0. These developments include: Jet
matching with Pythia parton showers for both Standard Model and Beyond
the Standard Model processes, decay chain functionality, decay width
calculation and decay simulation, process generation for the Grid, a
package for calculation of quarkonium amplitudes, calculation of
Matrix Element weights for experimental events, automatic dipole
subtraction for next-to-leading order calculations, and an interface
to FeynRules, a package for automatic calculation of Feynman rules and
model files from the Lagrangian of any New Physics model.
\end{abstract}

\maketitle

%%%%%%%%%%%%%%%%%%%%%%%%%%%%%%%%%%%%%%%%%%%%
%% MAINMATTER
%%%%%%%%%%%%%%%%%%%%%%%%%%%%%%%%%%%%%%%%%%%%

\section{Introduction}

With the imminent start of full-scale operation of the LHC experiment,
it has become increasingly important to have efficient and versatile
simulation tools, for New Physics signals as well as Standard Model
backgrounds. Much effort has gone into the development of new tools in
recent years, resulting in a number of packages addressing different
issues. These include a new generation of general-purpose tools
(Sherpa\cite{Gleisberg:2003xi}, Pythia8\cite{Sjostrand:2007gs},
Herwig++\cite{Gieseke:2003hm}), automatized matrix element generators
and event generators (AlpGen\cite{Mangano:2002ea},
CompHEP\cite{Boos:2004kh}/CalcHEP\cite{Pukhov:2004ca},
Helac\cite{Kanaki:2000ey}, Whizard\cite{Kilian:2007gr},
MadGraph/MadEvent\cite{Alwall:2007st}), and next-to-leading order
event generators (MCFM\cite{Campbell:2000bg},
MC@NLO\cite{Frixione:2006gn}).

MadGraph/MadEvent~\cite{Alwall:2007st} is a fully
automatized tool for generation of cross sections and unweighted
events for processes, both in the Standard Model and for models of New
Physics. MadGraph~\cite{Stelzer:1994ta} takes as input a process,
specified in a simple syntax, and a model definition. It is also
possible to specify multi-particle labels, the maximum order in the
different couplings (e.g.\ QCD and QED), and require or exclude
intermediate s-channel particles. MadGraph then produces all Feynman
diagrams for this process and all subprocesses, as well as its matrix
element expression in the form of a Fortran subroutine with calls to
the helicity amplitude library HELAS~\cite{Murayama:1992gi}.
MadEvent~\cite{Maltoni:2002qb} is used to perform the phase space
integration of the process (including any specified cuts) and produces
weighted and unweighted events, using a technique dubbed
``Single-diagram-enhanced multichannel integration'', which gives high
unweighting efficiencies also for multi-particle final states. This
technique has the additional advantage that it is trivially
parallelizable to run on multi-processor clusters. Events are output
to a text file following the Les Houches Accord for event
generation~\cite{Alwall:2006yp}, and interfaces to
Pythia~\cite{Sjostrand:2006za}, Herwig~\cite{Corcella:2002jc} and
PGS~\cite{PGS} allows for parton showering, hadronization and fast
detector simulation.

Event generation with MadGraph/MadEvent can be done online over the
Internet, on any of several dedicated computer clusters around the
world, using a simple but powerful web interface.  It is also possible
to download the source code to compile and run locally. Complete event
simulation using Pythia and PGS can be done online, directly or in
stages.

\section{Recent developments}

In version 4.0 of MadGraph/MadEvent~\cite{Alwall:2007st}, the
structure of MadGraph/MadEvent was modified to allow for implementation
of further models than the Standard Model, and new models were
implemented, including several simple extensions of the Standard Model
as well as the completely general Two Higgs Doublet Model~\cite{2HDM}
and the MSSM \cite{Cho:2006sx,MSSM}. Also a simple semi-automatized
framework for addition of user-defined extensions of the Standard
Model was provided. Automatization of the inclusion of new models in
MG/ME has now been further extended with the introduction of the
interface to the FeynRules package~\cite{Christensen:2008py}, see below.

Developments since version 4.0 include:
\begin{itemize}
\item Jet matching/merging between MG/ME and Pytha parton showers for
Standard Model~\cite{Alwall:2007fs} and Beyond the Standard Model processes
\cite{SUSYJets, deVisscher}
\item Decay chain functionality, which generates only the diagrams
consistent with a specified decay chain~\cite{decay-chains}
\item Particle decay width calculation and decay simulation directly
in MG/ME~\cite{decay}
\item Creation of MadEvent process packages suitable for event
generation on the Grid~\cite{GridWiki}
\item A repository of LHC events and Grid packages for important
Standard Model backgrounds to New Physics~\cite{LHCrepository}
\item MadOnium, calculation of quarkonium amplitudes in NRQCD~\cite{Artoisenet:2007qm}
\item MadWeight, a package for calculation of Matrix Element weights
of experimental events~\cite{MadWeight}
\item MadDipole, automatic dipole subtraction for next-to-leading
order real correction calculations~\cite{Frederix:2008hu}
\item Interface to FeynRules, a Mathematica package for automatic
calculation of Feynman rules and creation of event generator files
directly from the Lagrangian of any New Physics
model~\cite{FeynRulesInterface}
\end{itemize}

In the following we will discuss some of these developments in greater
detail.

\subsection{Jet matching}

The simulation of jet production from QCD bremsstrahlung emissions has
traditionally been done using Parton Shower (PS) Monte Carlo programs
such as Pythia and Herwig, which describe parton radiation as
successive parton emissions using the soft and collinear limit. This
description is formally correct only in the limit of soft and
collinear emissions, but has been shown to give a good description of
much data also relatively far away from this limit. However, for the
production of hard and widely separated jets in connection with heavy
particle production, this description breaks down due to the lack of
subleading terms and interference. For that case, it is necessary to
use the full tree-level amplitudes for the heavy particle production
plus additional partons. This description, however, diverges as
partons become soft or collinear. In order to
describe both these areas in phase space, the two descriptions must be
combined, without double counting or gaps between different
multiplicities. An additional physical requirement is that such a
procedure gives smooth distributions, and interpolates between the
parton shower description in the soft and collinear limits and the
matrix element description in the limit of hard and widely separated
partons. Several procedures have been proposed, including the
CKKW~\cite{Catani:2001cc,Krauss:2002up},
L\"onnblad~\cite{Lonnblad:2001iq} and Mangano~\cite{MLM}
schemes. These different procedures are in substantial agreement and
give consistent results at hadron
colliders~\cite{Alwall:2007fs,Mrenna:2003if}.

Two matching schemes are implemented in MadGraph/MadEvent interfaced to
Pythia. The first is a version of the Mangano scheme, but using $k_T$
jets instead of cone jets in the jet matching step. The details of
this scheme is described in sec.~2.4 of \cite{Alwall:2007fs}. This
scheme now works with both the ``old'' (virtuality-ordered) and the
``new'' ($p_T$-ordered) Pythia parton showers. The second is a new
scheme, which uses the feature of the $p_T$-ordered parton showers of
Pythia~6.4 that it stores the $k_T$ of the first emission in the
parton shower in a common block. An event is generated by MadEvent,
with QCD emission $\alpha_s$ values calculated as done in the parton
shower, using the $k_T$ values of the event clustered with the $k_T$
algorithm (as described in \cite{Alwall:2007fs}). The event is passed
to Pythia and showered using the $p_T$-ordered shower, which reports
the $k_T$ value of the first emission
$Q_\mathrm{max}^\mathrm{PS}$. This value is then compared to the
matching scale $Q_\mathrm{match}$, and the event is discarded if
$Q_\mathrm{max}^\mathrm{PS}>Q_\mathrm{match}$. The exception is for
the highest parton multiplicity sample, where the event is discarded
only if $Q_\mathrm{max}^\mathrm{PS}$ is higher than the lowest $k_T$
value in the matrix element multiparton event.  This matching scheme,
although simple, effectively mimics the workings of the $k_T$-jet
Mangano scheme. However, it more directly samples the Sudakov form
factor used in the shower. Furthermore, the treatment of the highest
multiplicity sample more closely mimics that used in the CKKW matching
scheme.

Both these schemes can be used in the Standard Model as well as in
processes involving New Physics. Jet matching in Beyond the Standard
Model processes was recently used in a studies of model-independent
gluino production at the Tevatron \cite{Alwall:2008ve, gluinolong},
where it was noted that inclusion of Matrix Element corrections is
crucial in the case of production of relatively light gluinos which
decay to a near-degenerate LSP, and hence producing only soft jets. In
this case, the only way to generate large missing transverse energy is
when the whole gluino pair center-of-mass system is boosted in the
transverse direction due to hard initial-state radiation, which is
well described only using jet matching.

Further studies of the effects of matching of jets in gluino and
squark production are underway, see \cite{SUSYJets, deVisscher}.

\subsection{Decay chains}

For precision studies of production and decay of new particles, many
features can be important. In particular, different types of New
Physics can often be distinguished using kinematical differences in decay
chains due to differences of spins of the intermediate particles. For
example, studies have shown that invariant masses between lepton pairs
\cite{Smillie:2005ar} and leptons and b jets \cite{Alves:2006df} can
be used to differentiate between Supersymmetry and Universal Extra
Dimensions. However, in order to correctly capture these differences
in simulations, it is necessary that the simulations keep the full
spin correlations in the decays. Also other effects can be
important, such as finite-width effects and interference effects from
non-resonant diagrams \cite{Berdine:2007uv}.

\begin{figure}
  \includegraphics[width=.9\columnwidth]{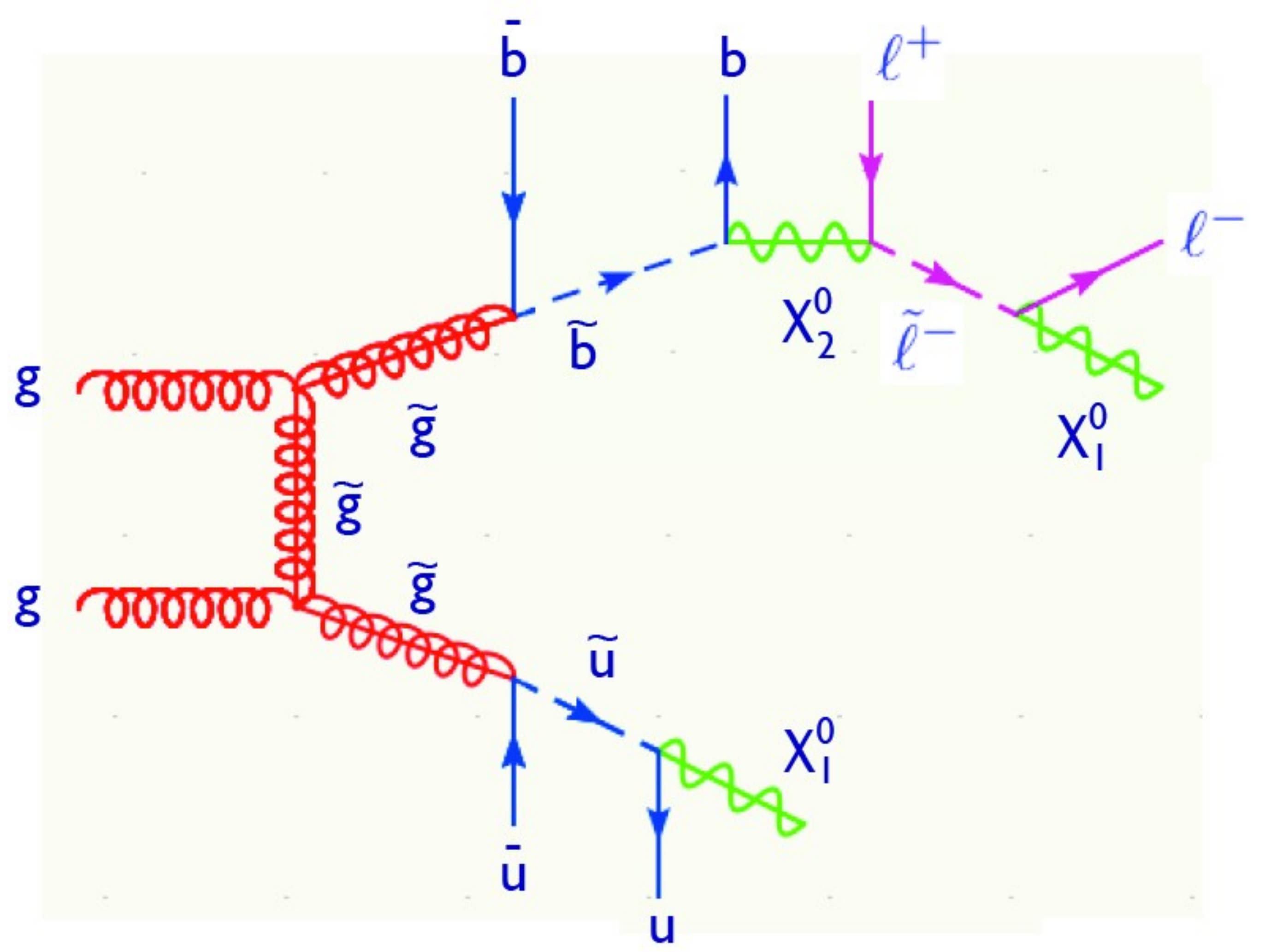}
  \caption{\label{fig:decaychain} An example of a decay chain that can
  be generated using the decay chain functionality of
  MadGraph/MadEvent.}
\end{figure}

Since MadGraph/MadEvent uses the full set of Feynman diagrams for any
process, it automatically keeps all spin correlations, as well as
finite-width effects and interference effects. However, running
MadGraph for many final state particles has been forbiddingly slow,
due to the large number of diagrams and diagram topologies. This
situation has now been considerably improved through the introduction
of the new decay chain functionality, which radically reduces diagram
generation times for decay chain-type processes, and allows for up to
10 external particles in a process. An example of such a process is
shown in Fig.~\ref{fig:decaychain}. This process can be specified
using the following syntax:
\begin{multline*}
\mathtt{gg>(go>u~(ul>u n1))(go>b~(b1>}\\
\mathtt{(b(n2>mu+(mul->mu- n1)))))}
\end{multline*}
(without line break.) Only diagrams conforming to this decay chain
specification are generated. In the subsequent MadEvent event
generation, all specified intermediate resonances are forced to be
onshell (i.e., have a virtuality at most a given number of widths off their
pole mass).

This technique for simulating decay chains has a number of advantages
over conventional simulation methods:

\begin{itemize}
\item The full matrix element is used, automatically keeping all spin
correlations in production and decay
\item Particle widths are correctly taken into account
\item $1\to N$ decays are possible while still keeping all
correlations
\item Non-resonant contributions can be included only where they are
relevant
\end{itemize}

It might be worthwhile to note that although this generation is
considerably faster than generation of the full multi-final state
matrix element including all interfering diagrams, it is still slower
than the conventional generation of the central process with
consecutive decays done in a second step. For scalar particle
production in particular, where no spin correlations are needed, it is
more efficient to use the MadGraph decay chain formalism up to the
production of the scalar particles, and then perform the decay of the
scalar particles using some decay tool such as
BRIDGE~\cite{Meade:2007js} or the MadGraph/MadEvent decay width
functionality, described below.

As a complement to the decay chain formalism described above, we have
also implemented the calculation of partial decay widths and
simulation of decay for unstable particles. This is done by specifying
the process according to the syntax
$$
\mathtt{A>BCD...}
$$
which indicates the decay of particle \texttt{A} to particles
\texttt{BCD...}. Multiparticle labels can be used as usual for the decay
products. When now MadEvent is run, it will present the partial decay
width instead of the cross section, and generate events corresponding
to the decay of \texttt{A} in its rest frame. These decays can
subsequently be boosted and combined with an event file describing
production of \texttt{A} particles.

\subsection{MadWeight}

MadWeight~\cite{MadWeight} is a package, built on top of MadGraph, to
find the Matrix Element weight of experimental events for a large set
of processes.  This procedure, also called the Matrix Element method,
is aimed at determining a set of free parameters $\bs \alpha$ of a
given theory from a data sample. It maximizes the information that can
be extracted from the detector by defining for each observed event
$\bs x$ a conditional weight $P(\bs x |\bs \alpha)$ that quantifies
the agreement between the theoretical model with parameters $\bs
\alpha$ and the experimental event $\bs x$.  In the definition of the
weights, the parton-level high-energy collsion is factorized from the
detector-level event, by introducing the parton-level configuration
$\bs y$ with a weight given by the squared matrix element $|M_{
\alpha}(\bs y)|^2$. The evolution of this parton-level configuration
$\bs y$ into a reconstructed event $\bs x$ in the detector is taken
into account by a transfer function $W(\bs y,
\bs x)$. As a result, the weight of a specific event $\bs x$ is of the
form
\begin{multline*}
P(\bs x |\bs \alpha)=
\frac{1}{\sigma_{ \alpha}} \int d \phi(\bs y) |M_{ \alpha (\bs y)}|^2
\times\\
dw_1 dw_2 f_1(w_1) f_2(w_2) W(\bs x,\bs y),
\end{multline*}
where $f_1(w_1)$ and $f_2(w_2)$ are the parton distribution functions.
The normalization by the total cross section $\sigma_{ \alpha}$
ensures that $P(\bs x |\bs \alpha)$ is a probability density. Once the
probability density $P(\bs x_i |\bs \alpha)$ has been computed for
each event $\bs x_i$, the most likely value for $\bs \alpha$ can be
obtained through a likelihood maximization method.

The numerical evaluation of the weights $P(\bs x|\bs\alpha)$ is in
general very complicated, since both the squared matrix element and
the transfer function are highly non-uniform over the phase space.
For an efficient Monte Carlo integration, phase-space points must be
generated according to the peak structure of the integrand. Therefore,
a specific phase-space generator, tuned to the shape of the transfer
function and the squared matrix element, is required. MadWeight
decomposes the phase space topology into substructures, where an
efficient integration can be achieved through suitable
parametrizations.

\begin{figure}
\includegraphics[width=1\columnwidth]{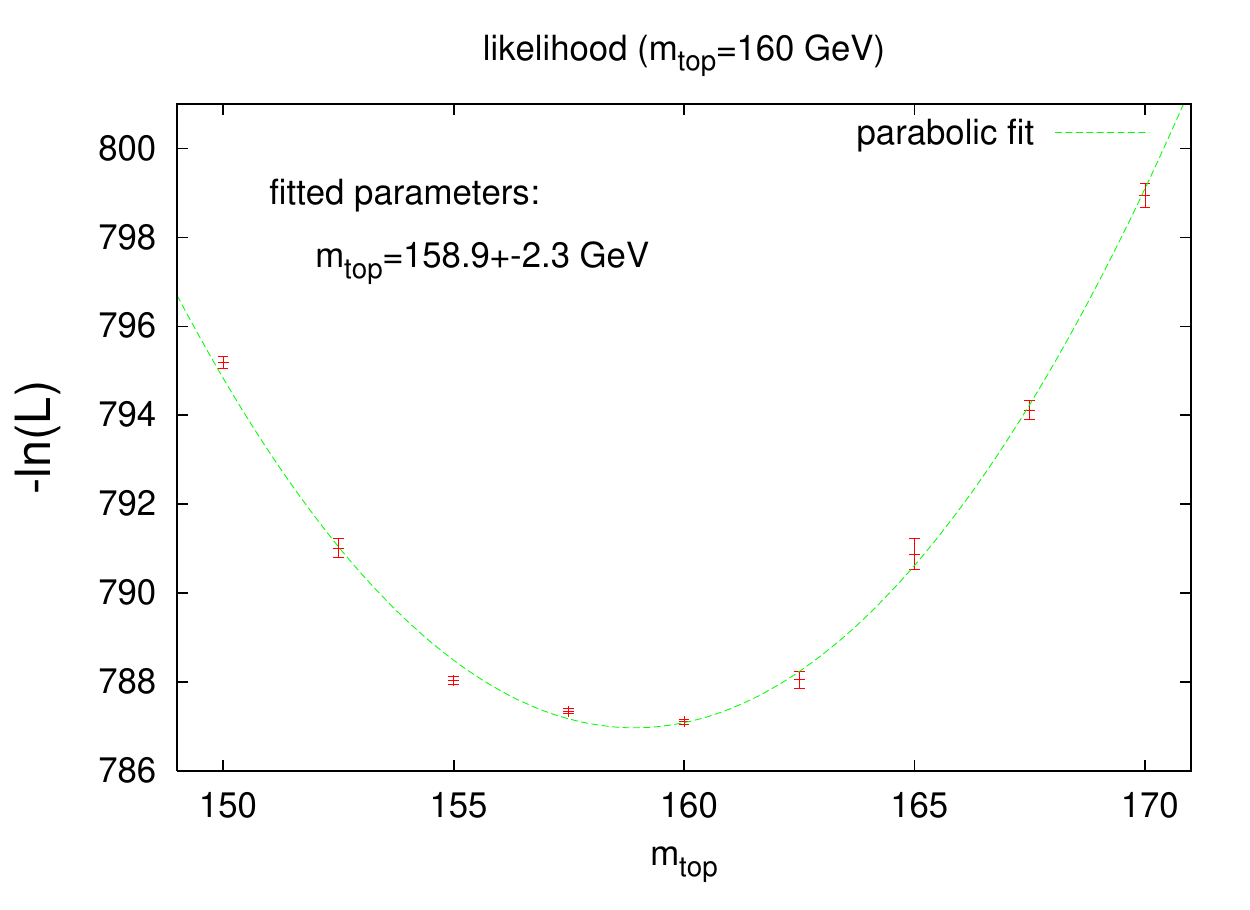}
\caption{\label{fig:MW:likelihood}Extraction of the top quark mass from a sample of 20 Monte Carlo $t \bar t$ events (semi-leptonic channel).}
\end{figure}

Fig~\ref{fig:MW:likelihood} provides an example of result that can be
obtained with MadWeight. The top-quark mass is extracted from a (MC)
sample of 20 $t\bar t$ events with one lepton in the final state, with
$m_t=160$ GeV in the simulation.  The weights are computed with
MadWeight and combined to get the likelihood displayed in
Fig.~\ref{fig:MW:likelihood}. The value of the reconstructed mass of
the top quark is $158.9 \pm 2.3$ GeV.

The types of processes that MadWeight can be used for include any
pair-production processes with decay chains. Also processes
with three or more particles produced in the central interaction can
be handled, such as pair production processes with radiation of an
additional parton. This provides for a large versatility and will allow
for Matrix Element methods to be used for many new types of processes.

\subsection{MadDipole}

MadDipole~\cite{Frederix:2008hu} is a tool for automatic calculation of
dipole-subtracted real corrections to next-to-leading order
calculation. Any next-to-leading (NLO)
computation in QCD includes the calculation of real contributions,
with the emission of an additional parton, and virtual contributions,
also called loop terms.  In general these suffer from soft and/or
collinear singularities, which only cancel after all contributions are
summed. Because the real contributions have one more particle in the
final state, numerical integrations over the phase spaces of the real
and virtual contributions need to be done independently, and hence the
divergencies must be extracted in a form where their cancellation is
explicit. One solution to remove the divergences due to the phase
space integration in the real contribution, is the subtraction scheme
proposed by Catani and Seymour~\cite{Catani:1996vz}.  In this method,
the Born contribution times a dipole function is added to both the
real and the virtual contributions, schematically
\begin{multline*}
\sigma^{\textrm{NLO}}=
\int_{m+1} \bigg[d\sigma^{\textrm{real}}-d\sigma^{\textrm{Born}}\otimes
D \bigg]+\\
\int_m \bigg[ \int_{\textrm{loop}}d\sigma^{\textrm{virt.}}
+ \int_1 d\sigma^{\textrm{Born}}\otimes D \bigg]_{\epsilon =0},
\end{multline*}
where $d\sigma^{\textrm{real}}$, $d\sigma^{\textrm{Born}}$ and
$d\sigma^{\textrm{virt.}}$ are the real, Born and virtual matrix
elements, respectively, and $D$ is the dipole function.  The dipole
functions are defined in such a way that the real term plus
subtraction term and the virtual plus subtraction term are
separately finite, and the contributions from the subtraction terms
cancel in the sum of the real and virtual contributions. 

MadDipole fully automatizes the calculation of the dipole subtraction
terms for massless and massive partons in the MadGraph/MadEvent
framework. The implementation is done in such a
way that the user only needs to specify the desired $(n + 1)$-particle
process, and MadDipole then returns a Fortran code for calculating all
dipoles, combined with possible Born processes which can lead to the
$(n + 1)$ process specified by the user. This works for processes in
any model. An implementation for automation of the calculation of
subtraction terms also for the virtual contributions is underway.

\subsection{Interface to FeynRules}

FeynRules~\cite{Christensen:2008py} is a new package based on
\emph{Mathematica}\textregistered\footnote{\emph{Mathematica} is a
registered trademark of Wolfram Research, Inc. } which takes a model
file with the Lagrangian as input and derives the interaction vertices
associated with this Lagrangian. The underlying algorithm, based on
canonical quantization formalism, is suitable not only for
renormalizable theories, but allows the derivation of the Feynman
rules in effective theories involving higher-dimensional operators as
well, which makes the package a useful tool for developing models
containing the SM as a low-energy effective theory.

\begin{figure}
  \includegraphics[width=.9\columnwidth]{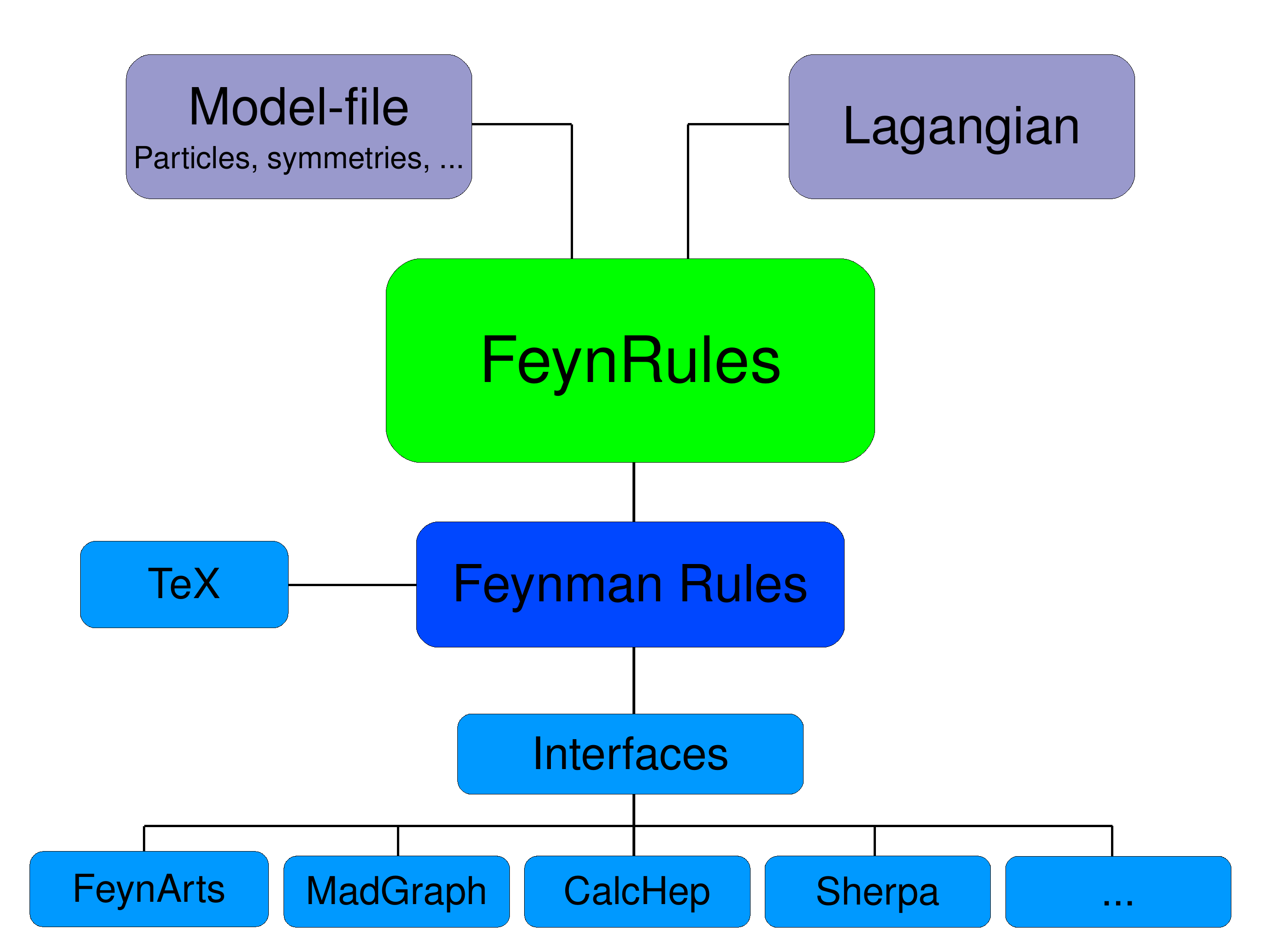}
  \caption{\label{fig:FeynRules} Logical organization of FeynRules.}
\end{figure}

The basic input a user provides when implementing his/her model into
FeynRules is the so called model file, a text file containing all the
properties of the model (particles, parameters, etc.), and the
Lagrangian written down using standard \emph{Mathematica} commands,
augmented by some new symbols like Dirac matrices, which are necessary
when writing down a Lagrangian.  The information contained in the
model file, together with the interaction vertices computed inside
\emph{Mathematica}, are stored in a generic format which is suitable
for any further processing of this information.  In a second step,
FeynRules can translate this generic model (with the vertices) into
the model format of choice and allows in this way to implement the new
model in any tool for which such an interface
exists. The organization of FeynRules is visualized in Fig.~\ref{fig:FeynRules}.

This approach allows FeynRules to go beyond previous packages
with similar functionality in several ways:
\begin{enumerate}
\item FeynRules is not tied to any existing Feynman diagram calculator.
\item The generic model format of FeynRules is suitable to be translated to any other format.
\item The user may choose his favorite Feynman diagram calculator, according to the strength and advantages of the latter.
\item The underlying Mathematica structure allows a
``theorist-friendly'' environment, which makes the package useful as a
``sandbox'' to develop new models. 
\end{enumerate}

Currently, interfaces have been written for CalcHEP/CompHEP,
FeynArts/FormCalc, MadGraph/MadEvent and Sherpa, and more
interfaces will be added in the future.

%Extensive tests have been done, both between the output of different
%Matrix Element generators using FeynRules-generated inputs (to check
%the interfaces) and between the results of FeynRules-generated inputs
%and previously implemented models. In particular the Standard Model
%and the MSSM has been carefully checked in this way, to ensure the
%correctness of the package.

Several models have been implemented in order to validate the code,
and extensive testing to check the interfaces is currently ongoing
both between the output of different Matrix Element generators using
the results of FeynRules-generated inputs and previously implemented
models. In particular the Standard Model and a higgsless three-site
model have been carefully checked in this way, to ensure the
correctness of the package, and further checks are ongoing for the
MSSM and extradimensional models.

\section{Conclusions}

With the recent developments of, and additions to,
MadGraph/MadEvent, the package has taken several further steps towards
a full-fledged simulation tool for physics beyond the Standard
Model. These developments include, in particular, the FeynRules
package which allows easy implementation of any new model into MG/ME;
calculation of decay widths and decays for any model, either using the
external BRIDGE package \cite{Meade:2007js} or MG/ME directly;
simulation of decay chain events, keeping full spin correlations and
finite width effects; matching of jets between MG/ME multiparton
generation and Pythia parton showers; and not least the analysis of
data events using matrix element techniques. MG/ME can now furthermore
be used directly as a tool for simulations of the real corrections to
next-to-leading order QCD calculations for multiparticle processes in
any model. Large-scale simulations can be done using the Grid with the
new grid-pack functionality, which is especially important for
Standard Model background event generation.

The MadGraph/MadEvent philosophy includes a strong emphasis of
building a user community, not only through fast response to questions
and bugs, but also by continuously develop the software and implement
requested features and functionalities. We also welcome and support
external efforts such as model implementations and development of
external packages and tools. And not least, we provide public clusters
with personal process databases, a Twiki site for communication, an
open CVS repository and, naturally, completely open source code.

%%%%%%%%%%%%%%%%%%%%%%%%%%%%%%%%%%%%%%%%%%%%%%%%
%% BACKMATTER
%%%%%%%%%%%%%%%%%%%%%%%%%%%%%%%%%%%%%%%%%%%%%%%%

\begin{theacknowledgments}
J.A.\ wishes to thank the organizers of SUSY08 for the invitation and
the generous support. J.A. is supported by the Swedish Research
Council.
\end{theacknowledgments}



\bibliographystyle{aipproc}   % if natbib is available

\bibliography{susy08-alwall}

\begin{thebibliography}{45}
\expandafter\ifx\csname natexlab\endcsname\relax\def\natexlab#1{#1}\fi
\providecommand{\enquote}[1]{``#1''}
\expandafter\ifx\csname url\endcsname\relax
  \def\url#1{\texttt{#1}}\fi
\expandafter\ifx\csname urlprefix\endcsname\relax\def\urlprefix{URL }\fi
\providecommand{\eprint}[2][]{\url{#2}}

\bibitem[Gleisberg et~al.(2004)]{Gleisberg:2003xi}
T.~Gleisberg, et~al., \emph{JHEP} \textbf{02}, 056 (2004),
  \eprint{hep-ph/0311263}.

\bibitem[Sjostrand et~al.(2008)]{Sjostrand:2007gs}
T.~Sjostrand, S.~Mrenna, and P.~Skands, \emph{Comput. Phys. Commun.}
  \textbf{178}, 852--867 (2008), \eprint{arXiv:0710.3820}.

\bibitem[Gieseke et~al.(2004)]{Gieseke:2003hm}
S.~Gieseke, A.~Ribon, M.~H. Seymour, P.~Stephens, and B.~Webber, \emph{JHEP}
  \textbf{02}, 005 (2004), \eprint{hep-ph/0311208}.

\bibitem[Mangano et~al.(2003)]{Mangano:2002ea}
M.~L. Mangano, M.~Moretti, F.~Piccinini, R.~Pittau, and A.~D. Polosa,
  \emph{JHEP} \textbf{07}, 001 (2003), \eprint{hep-ph/0206293}.

\bibitem[Boos et~al.(2004)]{Boos:2004kh}
E.~Boos, et~al., \emph{Nucl. Instrum. Meth.} \textbf{A534}, 250--259 (2004),
  \eprint{hep-ph/0403113}.

\bibitem[Pukhov(2004)]{Pukhov:2004ca}
A.~Pukhov  (2004), \eprint{hep-ph/0412191}.

\bibitem[Kanaki and Papadopoulos(2000)]{Kanaki:2000ey}
A.~Kanaki, and C.~G. Papadopoulos, \emph{Comput. Phys. Commun.} \textbf{132},
  306--315 (2000), \eprint{hep-ph/0002082}.

\bibitem[Kilian et~al.(2007)]{Kilian:2007gr}
W.~Kilian, T.~Ohl, and J.~Reuter  (2007), \eprint{arXiv:0708.4233}.

\bibitem[Alwall et~al.(2007{\natexlab{a}})]{Alwall:2007st}
J.~Alwall, et~al., \emph{JHEP} \textbf{09}, 028 (2007{\natexlab{a}}),
  \eprint{arXiv:0706.2334}.

\bibitem[Campbell and Ellis(2000)]{Campbell:2000bg}
J.~M. Campbell, and R.~K. Ellis, \emph{Phys. Rev.} \textbf{D62}, 114012 (2000),
  \eprint{hep-ph/0006304}.

\bibitem[Frixione and Webber(2006)]{Frixione:2006gn}
S.~Frixione, and B.~R. Webber  (2006), \eprint{hep-ph/0612272}.

\bibitem[Stelzer and Long(1994)]{Stelzer:1994ta}
T.~Stelzer, and W.~F. Long, \emph{Comput. Phys. Commun.} \textbf{81}, 357--371
  (1994), \eprint{hep-ph/9401258}.

\bibitem[Murayama et~al.(1992)]{Murayama:1992gi}
H.~Murayama, I.~Watanabe, and K.~Hagiwara  (1992), kEK-91-11.

\bibitem[Maltoni and Stelzer(2003)]{Maltoni:2002qb}
F.~Maltoni, and T.~Stelzer, \emph{JHEP} \textbf{02}, 027 (2003),
  \eprint{hep-ph/0208156}.

\bibitem[Alwall et~al.(2007{\natexlab{b}})]{Alwall:2006yp}
J.~Alwall, et~al., \emph{Comput. Phys. Commun.} \textbf{176}, 300--304
  (2007{\natexlab{b}}), \eprint{hep-ph/0609017}.

\bibitem[{Sj\"ostrand} et~al.(2006)]{Sjostrand:2006za}
T.~{Sj\"ostrand}, S.~Mrenna, and P.~Skands, \emph{JHEP} \textbf{05}, 026
  (2006), \eprint{hep-ph/0603175}.

\bibitem[Corcella et~al.(2002)]{Corcella:2002jc}
G.~Corcella, et~al.  (2002), \eprint{hep-ph/0210213}.

\bibitem[Conway et~al.(2006)]{PGS}
J.~Conway, et~al.  (2006), see
  \texttt{http://www.physics.ucdavis.edu/$\sim$conway/
  research/software/pgs/pgs4-general.htm}.

\bibitem[{Implementation by M.~Herquet and S.~de Visscher}(2006)]{2HDM}
{Implementation by M.~Herquet and S.~de Visscher}  (2006).

\bibitem[Cho et~al.(2006)]{Cho:2006sx}
G.~C. Cho, et~al., \emph{Phys. Rev.} \textbf{D73}, 054002 (2006),
  \eprint{hep-ph/0601063}.

\bibitem[{Implementation in MadEvent by J.~Alwall}(2006)]{MSSM}
{Implementation in MadEvent by J.~Alwall}  (2006).

\bibitem[Christensen and Duhr(2008)]{Christensen:2008py}
N.~D. Christensen, and C.~Duhr  (2008), \eprint{arXiv:0806.4194}.

\bibitem[Alwall et~al.(2008{\natexlab{a}})]{Alwall:2007fs}
J.~Alwall, et~al., \emph{Eur. Phys. J.} \textbf{C53}, 473--500
  (2008{\natexlab{a}}), \eprint{arXiv:0706.2569}.

\bibitem[Alwall et~al.(2008{\natexlab{b}})]{SUSYJets}
J.~Alwall, F.~Maltoni, and S.~{de Visscher}  (2008{\natexlab{b}}), {\it in
  progress}.

\bibitem[{de Visscher}(2008)]{deVisscher}
S.~{de Visscher}  (2008), in these proceedings.

\bibitem[{Implementation by J.~Alwall and T.~Stelzer}(2007)]{decay-chains}
{Implementation by J.~Alwall and T.~Stelzer}  (2007), use this publication as
  reference.

\bibitem[{Implementation by J.~Alwall}(2008)]{decay}
{Implementation by J.~Alwall}  (2008), use this publication as reference.

\bibitem[{The MadGraph Team}(2008{\natexlab{a}})]{GridWiki}
{The MadGraph Team}  (2008{\natexlab{a}}), see
  \texttt{http://cp3wks05.fynu.ucl.ac.be/twiki/
  bin/view/Library/GridDevelopment}.

\bibitem[{The MadGraph Team}(2008{\natexlab{b}})]{LHCrepository}
{The MadGraph Team}  (2008{\natexlab{b}}), see
  \texttt{http://cp3wks05.fynu.ucl.ac.be/twiki/
  bin/view/Library/MadGraphSamples}.

\bibitem[Artoisenet et~al.(2008{\natexlab{a}})]{Artoisenet:2007qm}
P.~Artoisenet, F.~Maltoni, and T.~Stelzer, \emph{JHEP} \textbf{02}, 102
  (2008{\natexlab{a}}), \eprint{arXiv:0712.2770}.

\bibitem[Artoisenet et~al.(2008{\natexlab{b}})]{MadWeight}
P.~Artoisenet, O.~Mattelaer, V.~Lemaitre, and F.~Maltoni  (2008{\natexlab{b}}),
  {\it in progress}.

\bibitem[Frederix et~al.(2008)]{Frederix:2008hu}
R.~Frederix, T.~Gehrmann, and N.~Greiner  (2008), \eprint{arXiv:0808.2128}.

\bibitem[{Implementation by C.~Duhr and M.~Herquet}(2008)]{FeynRulesInterface}
{Implementation by C.~Duhr and M.~Herquet}  (2008).

\bibitem[Catani et~al.(2001)]{Catani:2001cc}
S.~Catani, F.~Krauss, R.~Kuhn, and B.~R. Webber, \emph{JHEP} \textbf{11}, 063
  (2001), \eprint{hep-ph/0109231}.

\bibitem[Krauss(2002)]{Krauss:2002up}
F.~Krauss, \emph{JHEP} \textbf{08}, 015 (2002), \eprint{hep-ph/0205283}.

\bibitem[{L\"onnblad}(2002)]{Lonnblad:2001iq}
L.~{L\"onnblad}, \emph{JHEP} \textbf{05}, 046 (2002), \eprint{hep-ph/0112284}.

\bibitem[Mangano et~al.(2002)]{MLM}
M.~Mangano, et~al.  (2002), see \texttt{http://mlm.home.cern.ch/mlm/alpgen/}.

\bibitem[Mrenna and Richardson(2004)]{Mrenna:2003if}
S.~Mrenna, and P.~Richardson, \emph{JHEP} \textbf{05}, 040 (2004),
  \eprint{hep-ph/0312274}.

\bibitem[Alwall et~al.(2008{\natexlab{c}})]{Alwall:2008ve}
J.~Alwall, M.-P. Le, M.~Lisanti, and J.~G. Wacker  (2008{\natexlab{c}}),
  \eprint{arXiv:0803.0019}.

\bibitem[Alwall et~al.(2008{\natexlab{d}})]{gluinolong}
J.~Alwall, M.-P. Le, M.~Lisanti, and J.~G. Wacker  (2008{\natexlab{d}}), {\it
  in progress}.

\bibitem[Smillie and Webber(2005)]{Smillie:2005ar}
J.~M. Smillie, and B.~R. Webber, \emph{JHEP} \textbf{10}, 069 (2005),
  \eprint{hep-ph/0507170}.

\bibitem[Alves et~al.(2006)]{Alves:2006df}
A.~Alves, O.~Eboli, and T.~Plehn, \emph{Phys. Rev.} \textbf{D74}, 095010
  (2006), \eprint{hep-ph/0605067}.

\bibitem[Berdine et~al.(2007)]{Berdine:2007uv}
D.~Berdine, N.~Kauer, and D.~Rainwater, \emph{Phys. Rev. Lett.} \textbf{99},
  111601 (2007), \eprint{hep-ph/0703058}.

\bibitem[Meade and Reece(2007)]{Meade:2007js}
P.~Meade, and M.~Reece  (2007), \eprint{hep-ph/0703031}.

\bibitem[Catani and Seymour(1997)]{Catani:1996vz}
S.~Catani, and M.~H. Seymour, \emph{Nucl. Phys.} \textbf{B485}, 291--419
  (1997), \eprint{hep-ph/9605323}.

\end{thebibliography}

\IfFileExists{\jobname.bbl}{}
 {\typeout{}
  \typeout{******************************************}
  \typeout{** Please run "bibtex \jobname" to optain}
  \typeout{** the bibliography and then re-run LaTeX}
  \typeout{** twice to fix the references!}
  \typeout{******************************************}
  \typeout{}
 }

\end{document}